\documentclass{aa}

\usepackage{graphicx}	
\usepackage{amsmath}	
\usepackage{amssymb}	
\usepackage{xcolor}     
\usepackage{soul}   
\usepackage[draft]{hyperref}
\usepackage{txfonts}

\newcommand{\Msun}{M_{\odot}}
\newcommand{\MBH}{M_{\rm bh}}
\newcommand{\Rein}{\langle R_{\rm Ein}\rangle}
\newcommand{\DTmean}{\left< \delta t \right>_\lambda}
\newcommand{\DTmeanRef}{\left< \delta t \right>_{\lambda_0}}
\newcommand{\DTmeanLP}{\left< \delta t \right>_{\lambda,\text{geo}}}
\newcommand{\Rsrc}{R_\lambda}
\newcommand{\RsrcRef}{R_{\lambda_0}}
\newcommand{\LLE}{L/L_{\rm E}}

\newcommand{\zl}{z_{\rm L}}
\newcommand{\zs}{z_{\rm S}}
\newcommand{\dl}{D_{\rm L}}
\newcommand{\ds}{D_{\rm S}}
\newcommand{\dls}{D_{\rm LS}}
\newcommand{\Mstar}{\langle M_{\star}\rangle}
\newcommand{\DKS}{D_{\rm KS}}

\newcommand{\bd}{\begin{displaymath}}
\newcommand{\ed}{\end{displaymath}}
\newcommand{\be}{\begin{equation}}
\newcommand{\ee}{\end{equation}}
\newcommand{\beaa}{\begin{eqnarray*}}
\newcommand{\eeaa}{\end{eqnarray*}}
\newcommand{\bea}{\begin{eqnarray}}
\newcommand{\eea}{\end{eqnarray}}

\newcommand{\sref}[1]{Section~\ref{#1}}
\newcommand{\aref}[1]{Appendix~\ref{#1}}
\newcommand{\fref}[1]{Figure~\ref{#1}}

\newcommand{\tref}[1]{Table~\ref{#1}}
\newcommand{\eref}[1]{Equation~(\ref{#1})}

\begin{document} 

\title{Measuring accretion disk sizes of lensed quasars with microlensing time delay in multi-band light curves}
\titlerunning{Microlensing time delay in light curves}

\author{
J.~H.~H.~Chan \inst{\ref{epfl}}\and
K.~Rojas\inst{\ref{epfl}}\and
M.~Millon\inst{\ref{epfl}}\and
F.~Courbin\inst{\ref{epfl}}\and
V.~Bonvin\inst{\ref{epfl}} \and
G.~Jauffret \inst{\ref{epfl}}
}

\institute{
Institute of Physics, Laboratory of Astrophysique, \'Ecole Polytechnique F\'ed\'erale de Lausanne (EPFL), Observatoire de Sauverny, 1290 Versoix, Switzerland 
\label{epfl}
\goodbreak
}
\date{\today}

\abstract{
Time-delay cosmography in strongly lensed quasars offer an independent way of measuring the Hubble constant, $H_0$. However, it has been proposed that the combination of microlensing and source-size effects, also known as microlensing time delay can potentially increase the uncertainty in time-delay measurements as well as lead to a biased time delay. In this work, we first investigate how microlensing time delay changes with assumptions on the initial mass function (IMF) and find that the more massive microlenses produce the sharper distributions of microlensing time delays. We also find that the IMF has modest effect on the the magnification probability distributions. Second, we present a new method to measure the color-dependent source size in lensed quasars using the microlensing time delays inferred from multi-band light curves. In practice the relevant observable is the differential microlensing time delays between different bands. We show from simulation using the facility as Vera C. Rubin Observatory that if this differential time delay between bands can be measured with a precision of $0.1$ days in any given lensed image, the disk size can be recovered to within a factor of $2$. If four lensed images are used, our method is able to achieve an unbiased source measurement within error of the order of $20\%$, which is comparable with other techniques.
}

\maketitle

\section{Introduction} 
\label{sec:intro}

Strong gravitational lens effect occurs when light emitted from a background source is deflected due to the gravitational potential of a foreground object, resulting in multiple lensed images of the source. In particular, strongly lensed quasars provide us with a powerful tool to study galaxy evolution and to infer cosmological parameters. Careful use of the positions and magnification ratios of the lensed images allow to quantify the level of substructures in the lensing galaxy, which is directly related to how galaxies form and evolve and to the nature of the dark matter they contain \citep[e.g.][]{Harvey2020, SuyuEtal12,DalalKochanek02}. As quasars are variable sources, the time delays between the lensed images allow us to determine the time-delay distance in each system, which is inversely proportional to the Hubble constant $H_0$ \citep[e.g.,][]{Refsdal64, SuyuEtal10}. 
Over the years, time-delay cosmography with lensed quasars has achieved high precision on $H_0$ measurements, as carried out by the H0LiCOW\footnote{\href{https://shsuyu.github.io/H0LiCOW/site/}{http://h0licow.org}} collaboration \citep[$H_0$ Lenses in COSMOGRAIL\footnote{\href{http://www.cosmograil.org}{http://www.cosmograil.org}}'s Wellspring;][]{SuyuEtal17,WongEtal19}. Future progress will be reported in the new series papers TDCOSMO\footnote{\href{https://obswww.unige.ch/~lemon/tdcosmo-master/projects.html}{http://tdcosmo.org}} \citep[Time-Delay COSMOgraphy;][]{MillonEtal19}. 

One of the key ingredients for time-delay cosmography is to measure precise and accurate time delays. These are regularly obtained by the COSMOGRAIL and TDCOSMO collaborations from optical light curves for a rapidly increasing number of lensed quasars \citep[e.g.][]{MillonEtal20b,MillonEtal20,BonvinEtal18, BonvinEtal19b}. 
On the other hand, \citet[][hereafter TK18]{Tie&Kochanek18} proposed a possible source of bias in time-delay measurements, called microlensing time delay, also affecting lensed supernovae \citep{BonvinEtal19a}. This extra positive or negative delay may emerge when microlensing caustics unequally magnify different parts of the quasar accretion disk (or supernova shell). 
To estimate microlensing time delay in lensed quasar, we require high precision of lens modeling and the structure of accretion disk. 

The traditional techniques to study the inner components of quasars are microlensing and reverberation mapping. 
Both approaches are used to infer the structure of the quasar accretion disk, which is vital to understand the growth and evolution of Super Massive Black Holes (SMBH). 
In the first case, measuring the microlensing amplitude variability using light curves or single epoch spectra has served this purpose for decades \citep[e.g.,][]{Schechter&Wambsganss02, Kochanek04, MorganEtal10, DaiEtal10,MorganEtal18,CornachioneEtal20,RojasEtal14,RojasEtal20}. 
In the second case, reverberation mapping commonly monitors non-lensed quasars in multiple filters \citep[e.g.,][]{FausnaughEtal16,JiangEtal17,MuddEtal18,YuEtal18}.
Under the condition that the disk size is larger at longer wavelength, the time delays between the light curves in different bands provide a measurement of the size ($\tau\propto R\propto \lambda^{4/3}$): the larger the disk, the longer the delay. Similarly, we can apply the same methodology as in reverberation mapping to ``lensed quasars'', if light curves of the lensed images are available in several filters. Such data should be available in the near future from Vera C. Rubin Observatory \citep[hereafter Rubin;][]{IvezicEtal19}.

In this work, we investigate microlensing time delay and propose a new method to measure the disk size in lensed quasars using multi-filter light curves. This is complementary to the traditional microlensing work that only considers the amplitude of the microlensing and not the delay and distortion it introduces in the light curves. We use RX~J1131$-$1231 (hereafter J1131) as an example, as it has displays prominent microlensing events. The redshifts of lens and source are $\zl=0.295$ and $\zs=0.658$, respectively \citep{SluseEtal03}. Based on the $H_\beta$ line width from measured by \citet{SluseEtal03}, \citet{DaiEtal10} estimated the black hole mass to be $\MBH = (1.3\pm0.3)\times10^8\Msun$. \tref{tab:lensing_parameter} lists the lensing parameters used to generate magnification maps, corresponding to a macro lens model with a stellar mass fraction of $f_\star=0.2$ relative to a pure de Vaucouleurs model \citep{DaiEtal10,Tie&Kochanek18}.

The paper is organized as follows. The microlensing time delay with various initial mass functions is described in \sref{sec:microlensing_delay}. \sref{sec:disk_measurement} presents our method to estimate the source size using microlensing time delay and our conclusions are described in \sref{sec:conclusion}.

\section{Microlensing time delay in IMFs} 
\label{sec:microlensing_delay}

Reliable time-delay measurements between the strongly lensed images are key to obtain precise $H_0$ measurement. 
However, in presence of microlensing, a lensed image is affected by a excess (positive or negative) of time delay, due to the microlens magnification pattern. 
This additional delay does not necessarily cancel out when measuring cosmological time delays in several lensed quasars, as a result of the different excesses at multiple images. 
TK18 proposed microlensing time delay (MLTD, hereafter) for the first time and has computed statistically the microlensing time delays for two lensed quasars: J1131 and HE~0435$-$1223. 
The mean bias of the distribution of microlensing delays is of the order of a day, and for peculiar geometrical configurations this bias can reach several days. Other lensed quasars display nearly negligible microlensing time delay \citep{BonvinEtal18,BirrerEtal19,ChenEtal19,ShajibEtal19}. 
A method has been presented in \cite{ChenEtal18} to account for the effect of microlensing time delay in cosmological inference with strongly lensed quasars.

Estimating MLTD requires a microlensing magnification map and a model for the source accretion disk that is affected by microlensing. 
In this work, we follow the recipe provided by TK18 but we further investigate the impact of the adopted initial mass function (IMF) on magnification maps and MLTD.
In this section, we will describe how we generate magnification maps in \sref{subsec:mu_map}, and recap the basic of MLTD in \sref{subsec:microlensing_delay}.
We present the MLTD distributions with various IMFs in \sref{subsec:delay_distribution}.
\begin{figure*}[ht!]
\centering
\includegraphics[scale=0.5]{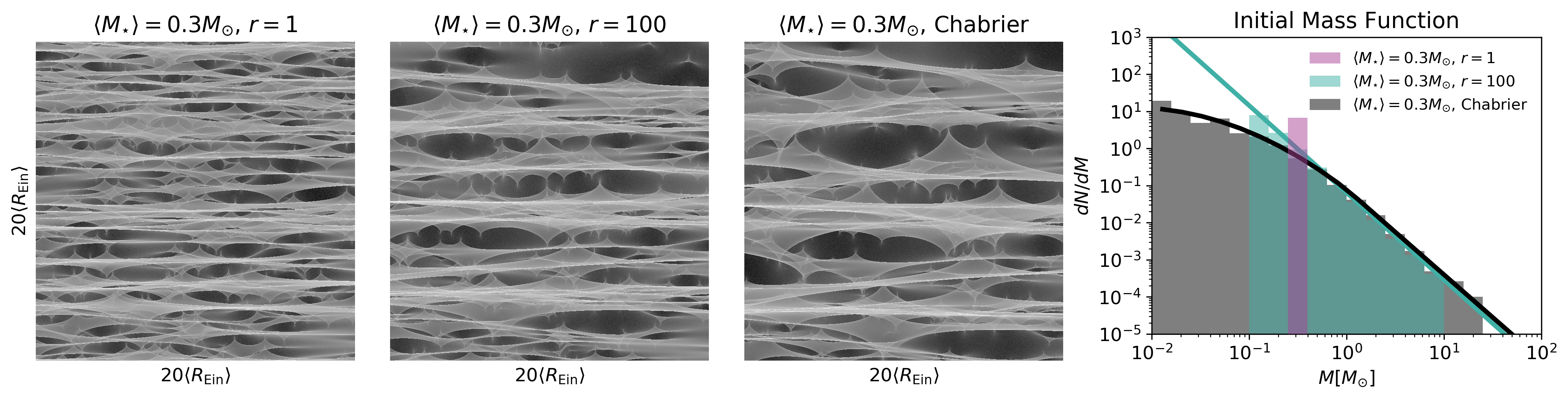}
\caption{Examples of magnification maps for image $A$ of RX~J1131$-$1231 on the left three panels, with the corresponding IMFs on the right panel.
All maps are $20\Rein$ on-a-side, corresponding to $8192^2$ pixels. 
The $r$ factor corresponds to the ratio between the upper to the lower bound of the mass interval, so that $r=1$ indicates single-mass IMFs. 
The magnification probability distributions are shown in \fref{fig:mu_distribution}.
}
\label{fig:mu_A}
\end{figure*}
\begin{figure*}[ht!]
\centering
\includegraphics[scale=0.5]{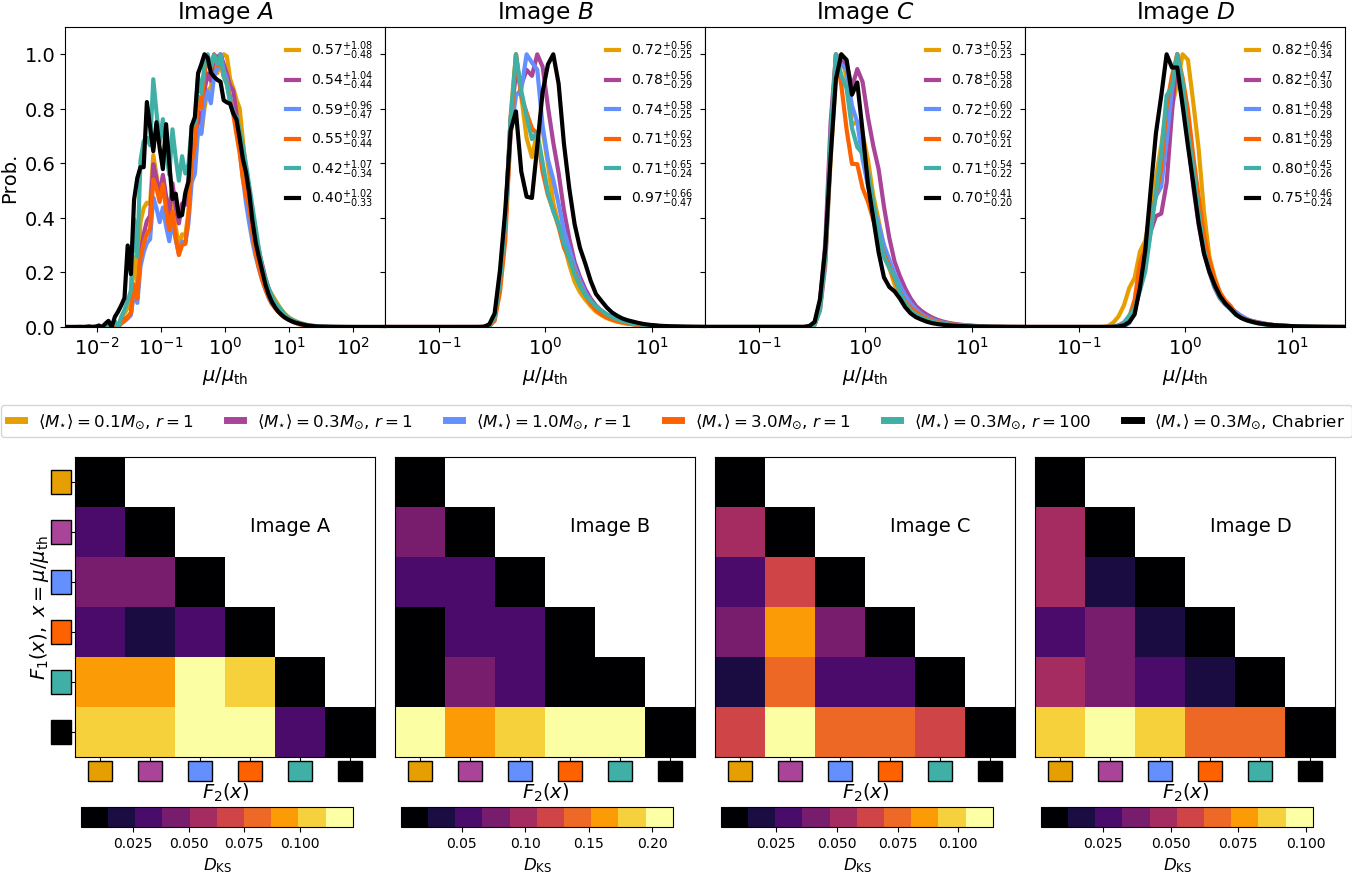}
\caption{
\textit{Top:} The magnification probability distribution of each lensed image, i.e. the histogram of the pixels in the maps (see \fref{fig:mu_A}).
\textit{ Bottom:} The Kolmogorov–Smirnov statistic of each distribution pair (see \eref{eqn:ks_stat}).
In each case the corresponding $\mu$ is indicated at the middle.
The Salpeter and Chabrier mass functions have slightly deviated distributions.
Small $\DKS$ shows that the probability distributions are insensitive, but not independent, to the chosen mean stellar mass or IMF. 
}
\label{fig:mu_distribution}
\end{figure*}

\subsection{Magnification maps}
\label{subsec:mu_map}

We generate the magnification maps using GPU-D \citep{Vernardos&Fluke14}, which includes a Graphics Processing Unit (GPU) implementation of the inverse ray–shooting technique \citep{KayserEtal86}. 
The default setup of the IMF is uniform with $1.0\Msun$.
In this work, we expand the range of mean microlens masses $\Mstar$ from $0.1\Msun$ to $3.0\Msun$.
Except for the uniform mass distribution, we adopt a Salpeter mass function with $\Mstar=0.3\Msun$ and the ratio between upper and lower bounds of the mass interval, $r=M_{\rm upper}/M_{\rm lower}=100$, which is favored by galaxy-scale lenses \citep[e.g.,][]{Kochanek04,OguriEtal14}. 
The Chabrier mass function with $\Mstar=0.3\Msun$ is also taken into account in this work \citep{Chabrier03}.

The microlensing parameters are taken from TK18, listed in \tref{tab:lensing_parameter}, i.e. the surface mass density for the macro model, $\kappa$ (lensing convergence), the shear of the macro model, $\gamma$, the fraction of the mass under the form of stars, $\kappa_{\star}/\kappa$, and the macro-magnification at the position of the quasar images, i.e.
\be
\mu_{\rm th}=\frac{1}{(1-\kappa^2)-\gamma^2}.
\ee
All magnification maps are $20\Rein$ on-a-side with 8192 pixels, which are sizable in this work, where
\begin{equation}
\begin{aligned}
\Rein = \sqrt{\frac{\ds\dls}{\dl}\frac{4G\Mstar}{c^2}} ,
\end{aligned}
\label{eqn:mean_rein}
\end{equation}
which depends on the angular diameter distances from the observer to the lens $\dl$, from the observer to the source $\ds$, and from the lens to the source $\dls$. 
For J1131, $\Rein=17.68~\text{light-day}\sqrt{\Mstar/\Msun}$. 
\begin{table}[t!]
    \begin{center}
\begin{tabular}{ccccr}
\hline
\hline
Image & $\kappa$ & $\gamma$ & $\kappa_{\star}/\kappa$ & $\mu_{\rm th}$\\
\hline
$A$ & $  0.618$ & $  0.412$ & $ 0.0667$ &$-41.982$ \\
$B$ & $  0.581$ & $  0.367$ & $ 0.0597$ &$ 24.467$ \\
$C$ & $  0.595$ & $  0.346$ & $ 0.0622$ &$ 22.569$ \\
$D$ & $  1.041$ & $  0.631$ & $ 0.1590$ &$ -2.522$ \\
\hline
\end{tabular}
\end{center}

    \caption{
    Microlensing model parameters of J1131: $\kappa$, $\gamma$, $\kappa_{\star}/\kappa$, and $\mu_{\rm th}$ at each lensed image position from TK18.
    }
    \label{tab:lensing_parameter}
\end{table}

As an example, we show in \fref{fig:mu_A} the magnification maps for image $A$ with the uniform, Salpeter, and Chabrier IMFs.
For $r=1$ this is equivalent to a delta function, corresponding to the uniform mass function.
The magnification probability distributions are shown on the top panels of \fref{fig:mu_distribution}.
In order to measure the similarities between different distributions, we adopt the Kolmogorov–Smirnov statistic for each pair of cumulative distribution functions $F_1(x)$ and $F_2(x)$, expressed as
\be
\DKS=\sup_x \left|F_1(x)-F_2(x)\right|,
\label{eqn:ks_stat}
\ee
where $\sup_x$ is the supremum of the set of distances.
$\DKS$ converges to $0$ when two distributions are the same.
The results are shown on the bottom panels of \fref{fig:mu_distribution}.
Given a sample size of $100$, the condition at the confidence level $95\%$ to reject that two distributions are the same is $\DKS\gtrsim0.2$.
Since the size of magnification map is scaled by mean stellar mass, once the shape of IMF is fixed, alternation of $\Mstar$ does not vary the map.
For the Salpeter and Chabrier mass functions, we note that the distributions are slightly deviated due to some larger caustics.
Although the details of the microlensing pattern do change, we find that the magnification probability distributions are insensitive, but not independent, to the choice of the IMFs. 
This result is in line with previous work \citep[e.g.][]{Wambsganss92,Lewis&Irwin95,Wyithe&Turner01,SchechterEtal04,Vernardos&Fluke13,MediavillaEtal15,Jimnez-Vicente&Mediavilla19}.

\subsection{Microlensing time delay}
\label{subsec:microlensing_delay}
\begin{figure*}[t!]
\centering
\includegraphics[scale=0.6]{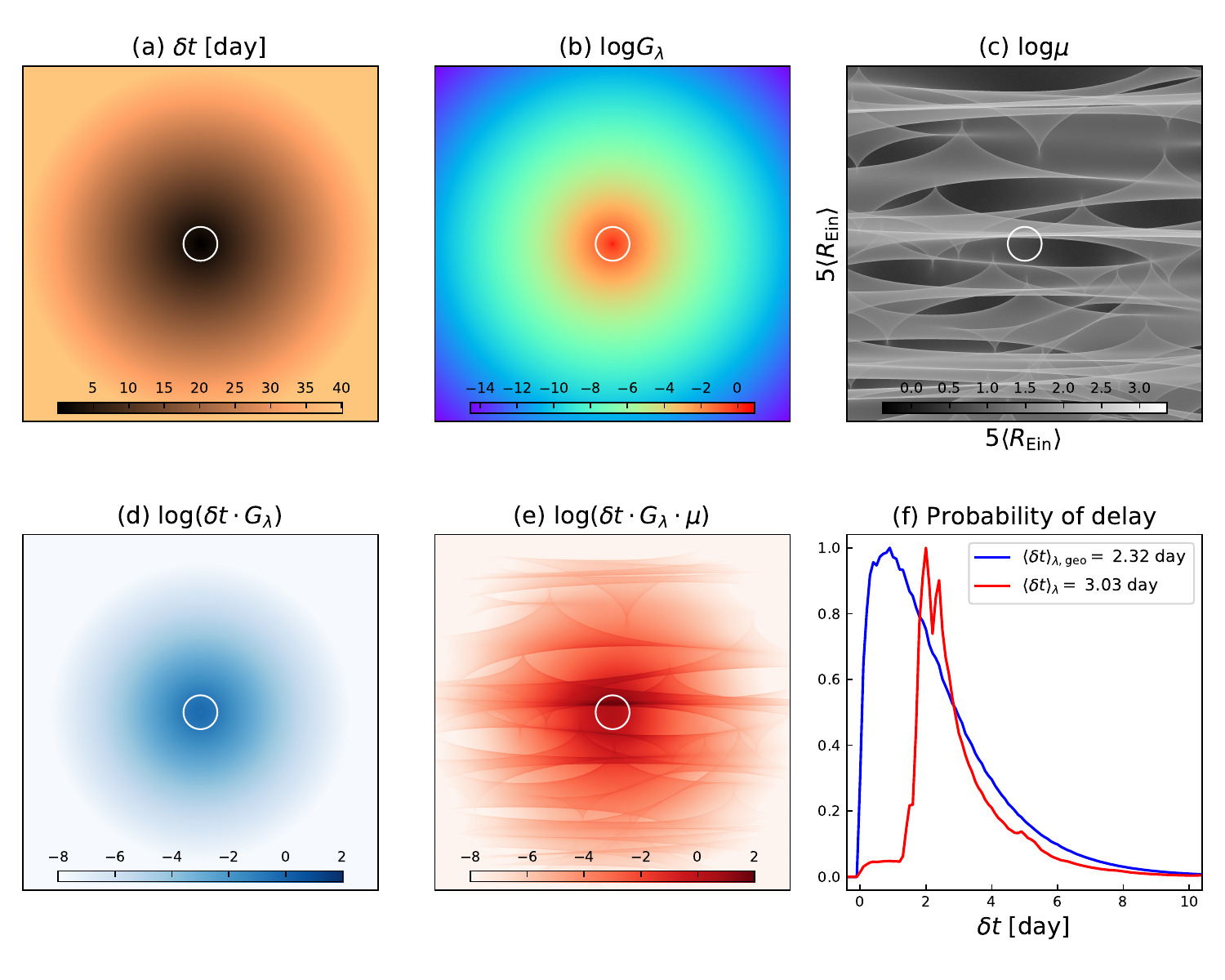}
\vskip -10pt
\caption{
Illustration of the distortion of the probability distribution of time delays with and without microlensing effect: a) the geometrical time lag of the disk which is simply $\delta t=(1+\zs)\sqrt{x^2+y^2}/c$, when the disk is seen face-on, b) the luminosity profile $G_{\lambda}$ with $\Rsrc=0.277$~light-day, c) a zoom-in of $5\Rein$ magnification pattern using $\Mstar=0.3\Msun,\ r=100$. The (d) and (e) panels show the time lag weighted by $G_{\lambda}$ and $G_{\lambda}\cdot\mu$, respectively. Panel (f) compares the net effect of microlensing on the time delay distribution we would obtain without microlensing.
The white circle in each color map is labeled with the effective radius $5.04(1+\zs)\Rsrc=2.315$~light-day.
}
\label{fig:delay_ml}
\end{figure*}

We now recall the basic of MLTD and describe the model used to generate simulations. 
For the lensed source, we consider a non-relativistic thin-disk model, emitting black-body radiation \citep{Shakura&Sunyaev73}, where the central source is assumed to be a ``lamp post'' located closely above the black hole \citep{CackettEtal07,KrolikEtal91,WandersEtal97,CollierEtal98,StarkeyEtal16}.
The temperature profile on a disk follows a simple form of $T(R)\propto R^{-3/4}$.
Although some observations support the shallower profile, like a slim disk \citep{AbramowiczEtal88,Cornachione&Morgan20}, a simple model should be allowed since MLTD is subtle and created by the variable flux \citep{Tie&Kochanek18}.
Under the thin-disk assumption, the disk radius where the disk temperature matches the photon wavelength ($kT=hc/\lambda$) can be expressed as
\be
\Rsrc
=3.745\,
\left(\frac{\lambda}{\rm \mu m} \right)^{4/3}
\left(\frac{\MBH}{10^9\Msun} \right)^{2/3}
\left(\frac{L}{\eta L_{\rm E}} \right)^{1/3} \text{light-days},
\label{eqn:source_size}
\ee
where $\LLE$ is the luminosity in unit of the Eddington luminosity, $L_{\rm E}$, and $\eta$ is the accretion efficiency \citep[e.g.][]{MorganEtal10}. Ignoring the inner edge of the ``face-on'' disk, in the simple lamp post model, the average MLTD can be derived using Eq.~10 of \citet{Tie&Kochanek18}, or Eq.~6 of \citet{ChanEtal19}, reproduced here for convenience:
\be
\DTmean
= \frac{1+z}{c} 
\frac{\int G_\lambda(x,y)\mu(x,y)\sqrt{x^2+y^2}dxdy}
     {\int G_\lambda(x,y)\mu(x, y)dxdy},
\label{eqn:mean_delay}
\ee
where $\mu(x, y)$ is the magnification map projected on the source plane with the coordinates $x,y$, and $G_\lambda(\xi)$ is the first derivative of the luminosity profile of the disk which can be expressed as
\be
G_{\lambda}(\xi)=\frac{\xi\exp(\xi)}{\left[\exp(\xi)-1\right]^2},
\ee
where $\xi=(\sqrt{x^2+y^2}/\Rsrc)^{3/4}$ \citep[see][for a detailed explanation of the coordinate system]{Tie&Kochanek18,ChanEtal19}. For a given geometrical configuration and accretion disk model, we can thus compute the mean excess of microlensing time delay $\DTmean$ for a given source position and magnification pattern. In absence of microlensing, all the lensed images share the same amount of lag $\DTmeanLP=5.04(1+\zs)\Rsrc/c$, which is called \textit{geometric delay} or \textit{lamp-post delay}. 
In \fref{fig:delay_ml}, we show how $\DTmean$ changes when the microlensing effect is added. In each panel we highlight the effective radius $5.04(1+\zs)\Rsrc=c\cdot\DTmeanLP=2.315\text{~light-day}$ as a white circle. The overall results are summarized in the lower right panel of the figure, which clearly shows how microlensing distorts the probability distribution of time delays.

\begin{figure*}[h!]
\centering
\includegraphics[scale=0.7]{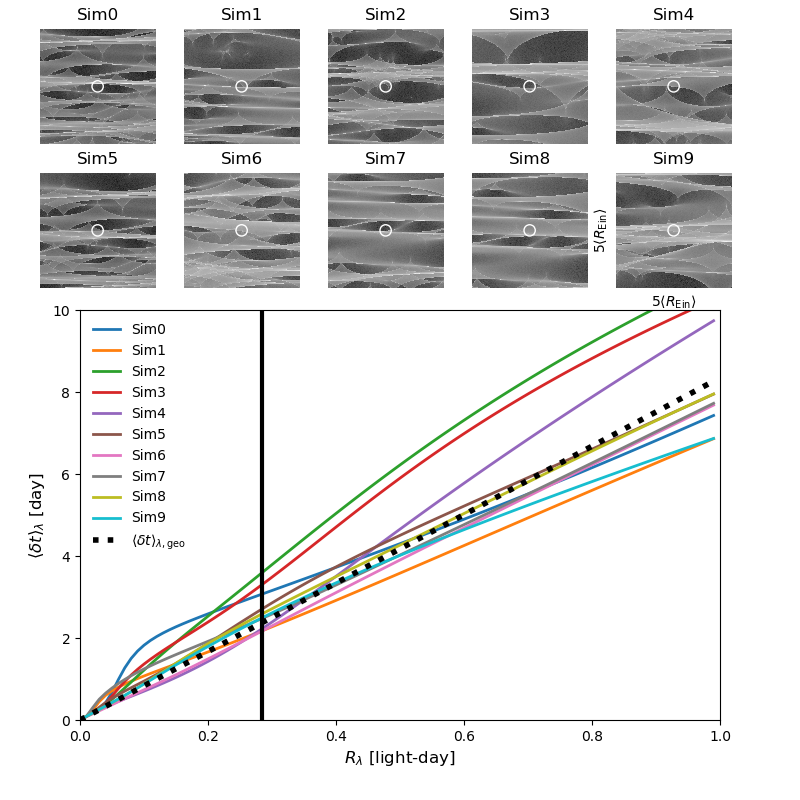}
\vskip -15pt
\caption{
{\it Top:} Examples of magnification patterns spanning each $5\Rein$, assuming the IMF $\Mstar=0.3\Msun$ and $r=100$. {\it Bottom:} Source size $\Rsrc$ and mean delay $\DTmean$ for each realisation of the magnification pattern. 
The dotted line in black corresponds to the relation without microlensing, i.e. the geometric delay.
The white circles on the top panels represent the effective radius $5.04(1+\zs)\Rsrc=2.315$~light-day, and the vertical black line on the bottom panel indicated $\Rsrc=0.277$~light-day (see text). 
}
\label{fig:sim}
\end{figure*}
\begin{figure*}[ht!]
\centering
\includegraphics[scale=0.5]{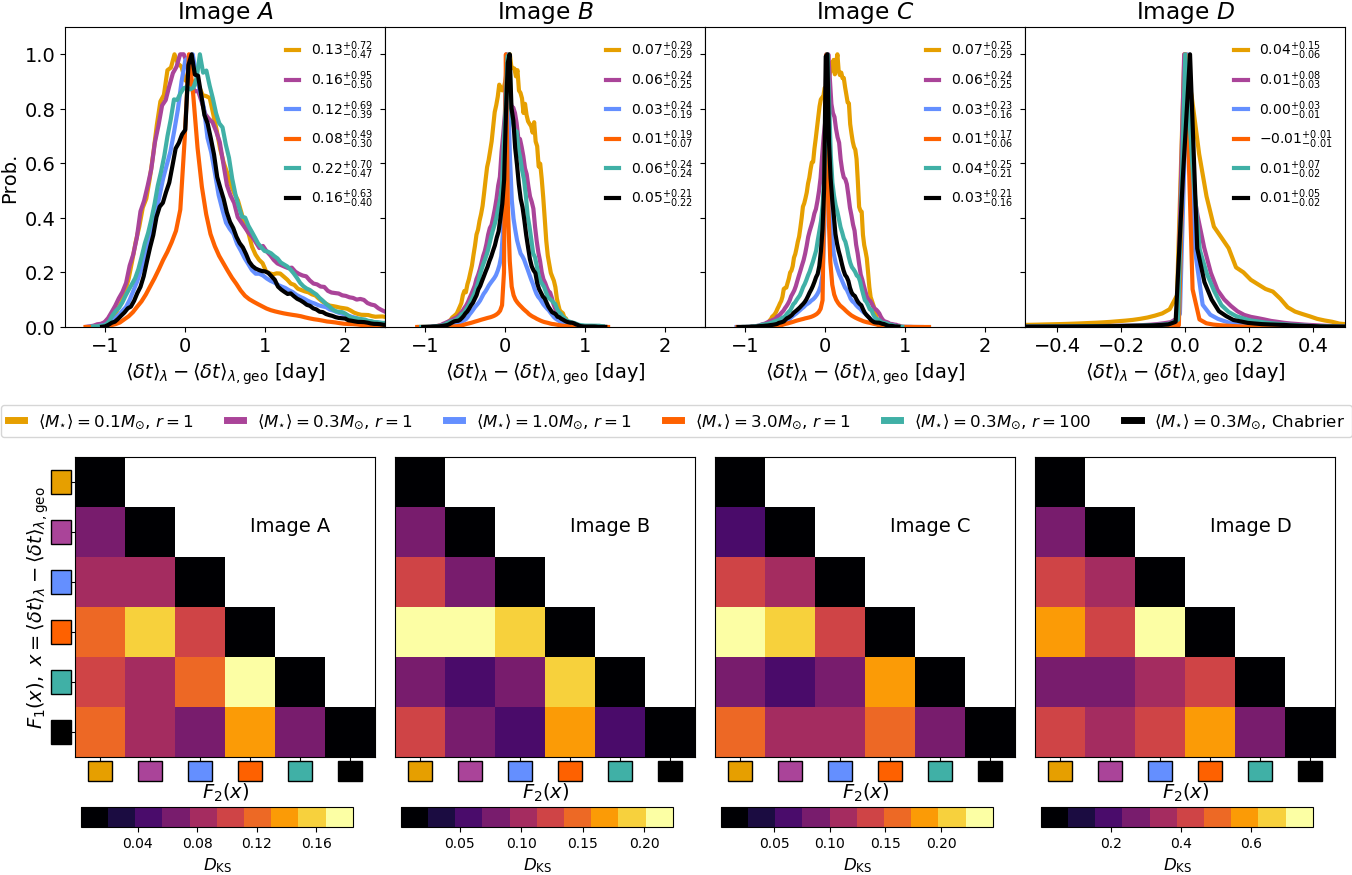}
\caption{
{\textit Top:} The probability distribution of microlensing time delay of each lensed image.
{\textit Bottom:} The Kolmogorov–Smirnov statistic of each distribution pair (see \eref{eqn:ks_stat}).
In each case the corresponding mean time delay is indicated at the middle.
}
\label{fig:delay_distribution}
\end{figure*}

The mean delay $\DTmean$ changes by varying the magnification pattern.
In \fref{fig:sim}, we present $\DTmean$ as a function of disk size $\Rsrc$, when a source is placed in $10$ random positions on a magnification map, corresponding $10$ sub-magnification patterns with a size of $5\Rein$, labelled from Sim0 to Sim9. The purely geometrical delay $\DTmeanLP$ is labeled as a black dot line, i.e. the delay in absence of microlensing. 
In the figure, we again take $\Rsrc=0.277$~light-day as fiducial value, highlighted by the vertical black line. As in \fref{fig:sim}, the effective radius $5.04(1+\zs)\Rsrc=2.315$~light-day is labeled as a white circle. 
When the inner circle is magnified, $\DTmean$ is shorter, as shown in Sim1 and Sim6. On the contrary, when the inner circle is demagnified, $\DTmean$ is longer, as shown in Sim2 and Sim3.

\subsection{The effect of IMF on microlensing time delay distribution}
\label{subsec:delay_distribution}

It is not possible to know the exact microlens pattern and we need to proceed in a statistical way. 
For a given geometrical configuration of the accretion disk model, we can then compute the mean excess of microlensing time delay $\DTmean$ at any given source position, as described in \sref{subsec:microlensing_delay}. 
By varying the magnification pattern, we can draw the probability distribution of microlensing time delay $\DTmean$ for each lensed image.
Practically, we can convolve a magnification map and disk light distribution using \eref{eqn:mean_delay}, and we then obtain the mean delay $\DTmean$ map, given a source at each pixel of the magnification map.
Although the magnification probability distribution is insensitive to IMFs as described in \sref{subsec:mu_map}, it is natural to ask whether the IMFs have an impact on the $\DTmean$ distribution. 
In the following section, we choose as a reference the filter used currently for cosmological time delay by the TDCOSMO collaboration \cite[e.g.][]{MillonEtal19}, i.e. the $R_c$ filter. For J1131, this corresponds to the observed wavelength $\lambda_{\rm obs} = (1+\zs)\times \lambda = 6517.25$~\AA.
Given an Eddington ratio of $\LLE=0.1$ and radiative efficiency of $\eta=0.1$, we obtain the disk size $\Rsrc=0.277$~light-day.

In \fref{fig:delay_distribution}, we present the microlensing time delay distributions of $\DTmean-\DTmeanLP$ for different IMFs on the top panels. 
This quantity is the excess of microlensing time delay with respect to the purely geometric time delay, since $\DTmeanLP$ is in absence when measuring cosmological time delay. 
For smaller mean stellar masses, we notice that the distribution becomes wider and has slightly larger mean.
Since microlensing time delay is estimated by the convolution of magnification map and disk light profile, the $\DTmean$ maps are similar to but smoother than the magnification map \citep[See Figures 2-5 in][]{Tie&Kochanek18}.
In other words, more small caustics located across the disk leads to a higher probability to produce small delays and thus the probability distribution becomes wider. 
Furthermore, since we can scale the size unit to $\Rein\propto\sqrt{\Mstar}$, smaller mean mass correlates with larger source size, resulting not only to the width but also to the offset of MLTD distribution.
In this work, we extend $\Mstar$ ranging by a factor of $30$, corresponding to $\Rsrc$ ranging by a factor of $5.5$. 
This offset can be evident at small mean masses or larger source sizes, which agrees with the result in TK18.
Same as \fref{fig:mu_distribution}, the Kolmogorov–Smirnov statistics are shown on the bottom panel.
We note that $\DKS$ of MLTD are generally larger than those of $\mu$.
This shows that the MLTD distributions are more sensitive to the mean stellar masses, resulting from the equivalent changes of source size, in particular for the sharper distributions, happening when $\Mstar=3.0\Msun$ and also in the image $D$.
However, a sharper distribution also indicates that the MLTD has less impact on the cosmological time delays.

\begin{figure*}[t!]
\centering
\includegraphics[scale=0.75]{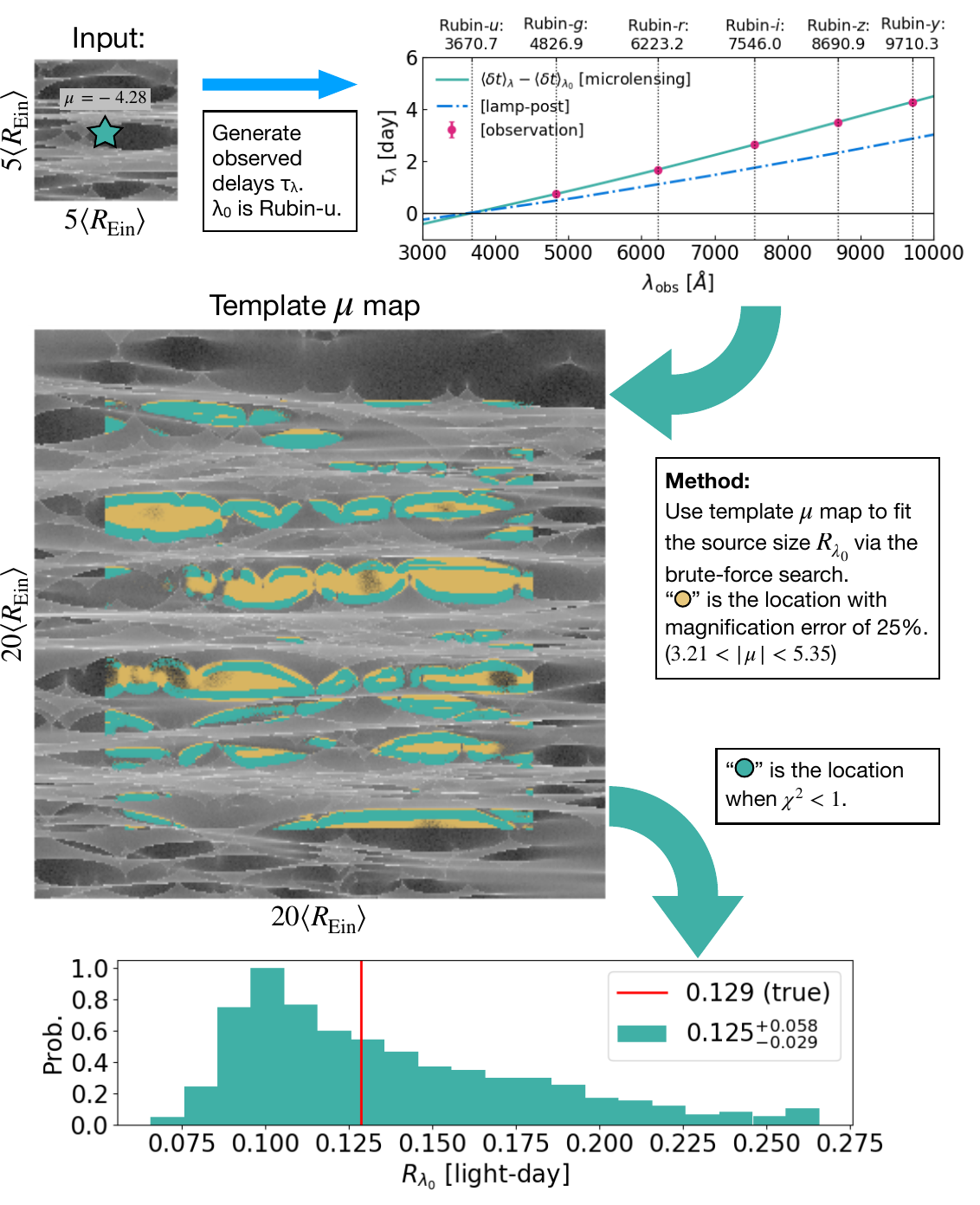}
\caption{
Flow chart of our procedure to simulate multi-band microlensing curve and to analyse them. The top row shows how we simulate our ``observed data'' given a fiducial source, magnification pattern and microlensing magnification, $\mu = -4.28$. The negative magnification indicates that the image is at the saddle point. The ``template magnification map'' used to analyse the data is shown in the middle panel (see text), where preselected locations used in fitting the source size are shown in yellow. These correspond to areas where the microlensing magnification matches the fiducial one with $\le25\%$ error. We finally keep only the regions where fitting the source size gives $\chi^2<1$. These are shown in green and are used to produce the distribution of measured source size $\RsrcRef$ on the bottom row. We give for reference the true (fiducial) size as a red vertical line.
 }
\label{fig:method}
\end{figure*}

\section{Disk measurement via microlensing time delay}
\label{sec:disk_measurement}

Reverberation mapping of continuum light curves is of big importance to measure the physical size of quasar accretion disks \citep{MuddEtal18,YuEtal18}. Measuring the relative lags between the multi-band light curves allows us to derive the color-dependent disk size, given a combination of an accretion disk model in the context of the lamp-post model for the intrinsic variations of the quasar. In principle, this method can be applied to lensed quasars, especially the quadruply-imaged ones that display four realisations of microlensing maps. However, the light curves of the lensed images are distorted by the microlensing with respect to each other, hence affecting the distributions for $\DTmean$ and $\Rsrc$, as shown in \fref{fig:sim}. 

In this section, we propose in the following a new method to estimate the disk size using microlensing time delay given multi-band light curves. We see this method both as a way to study quasar physics at very small scales (light days) and as a way to mitigate the impact of microlensing time delay when measuring $H_0$. We illustrate our method in \fref{fig:method} and give the technical details in \sref{subsec:method} along with our main findings in \sref{subsec:result}.

\begin{figure*}[pt!]
\centering
\includegraphics[scale=0.44]{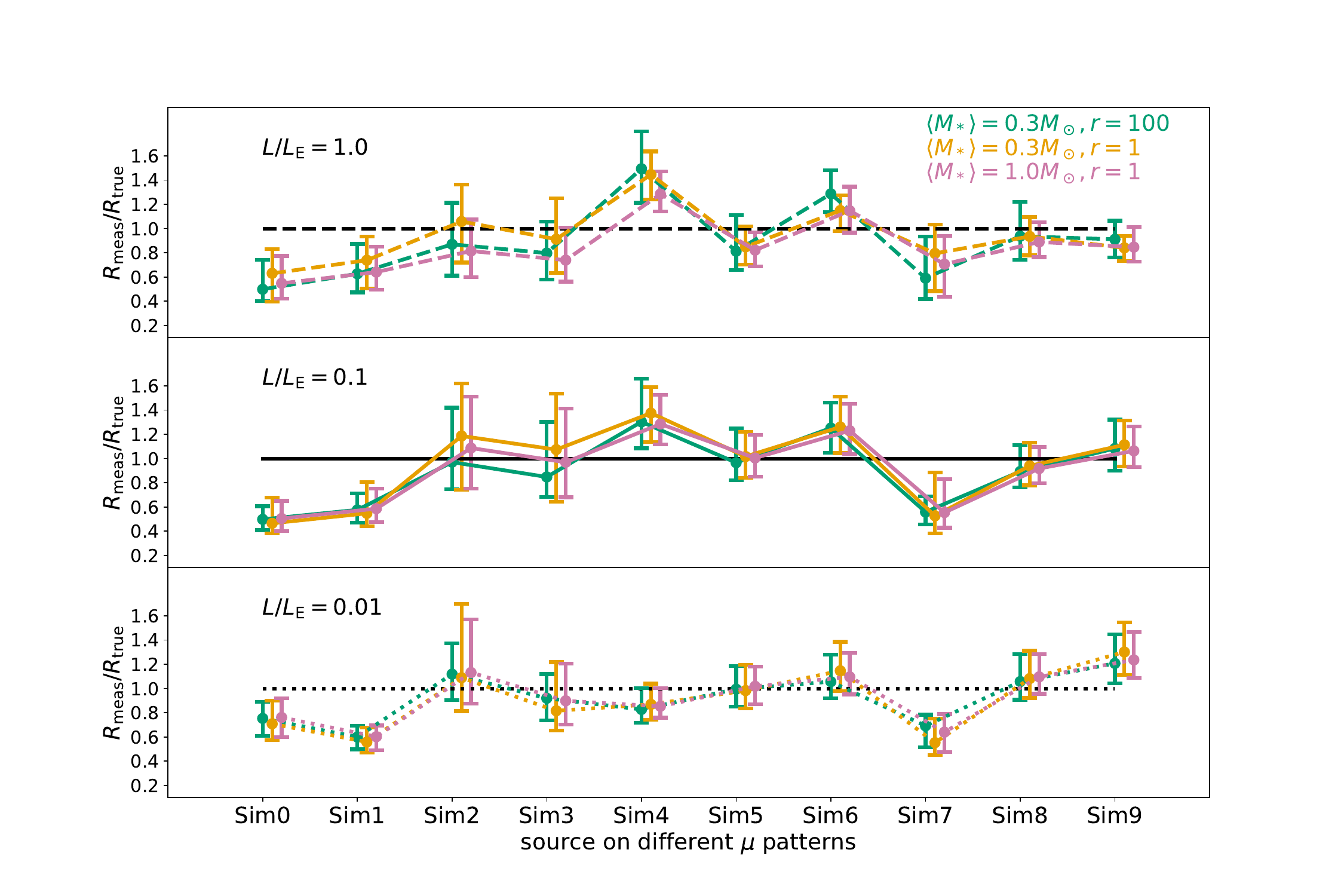}
\caption{Ratio between source size measurement and true size for 10 simulations of microcaustic networks with lensing properties corresponding to quasar image $A$ of J1131. Using for three assumed luminosities $\LLE=1.0$, $0.1$, and $0.01$, the source size corresponds to $\RsrcRef=0.277$, $0.129$, and $0.060$~light-day, respectively (see Eq.~\ref{eqn:mean_rein}). The unbiased measurements are labeled as black lines as reference. Each time, three typical IMFs are implemented and labeled on the top right corner. 
Clearly the results are insensitive to the choice of the lens galaxy IMF.
}
\label{fig:r0_imf}
%
%
\centering
\includegraphics[scale=0.44]{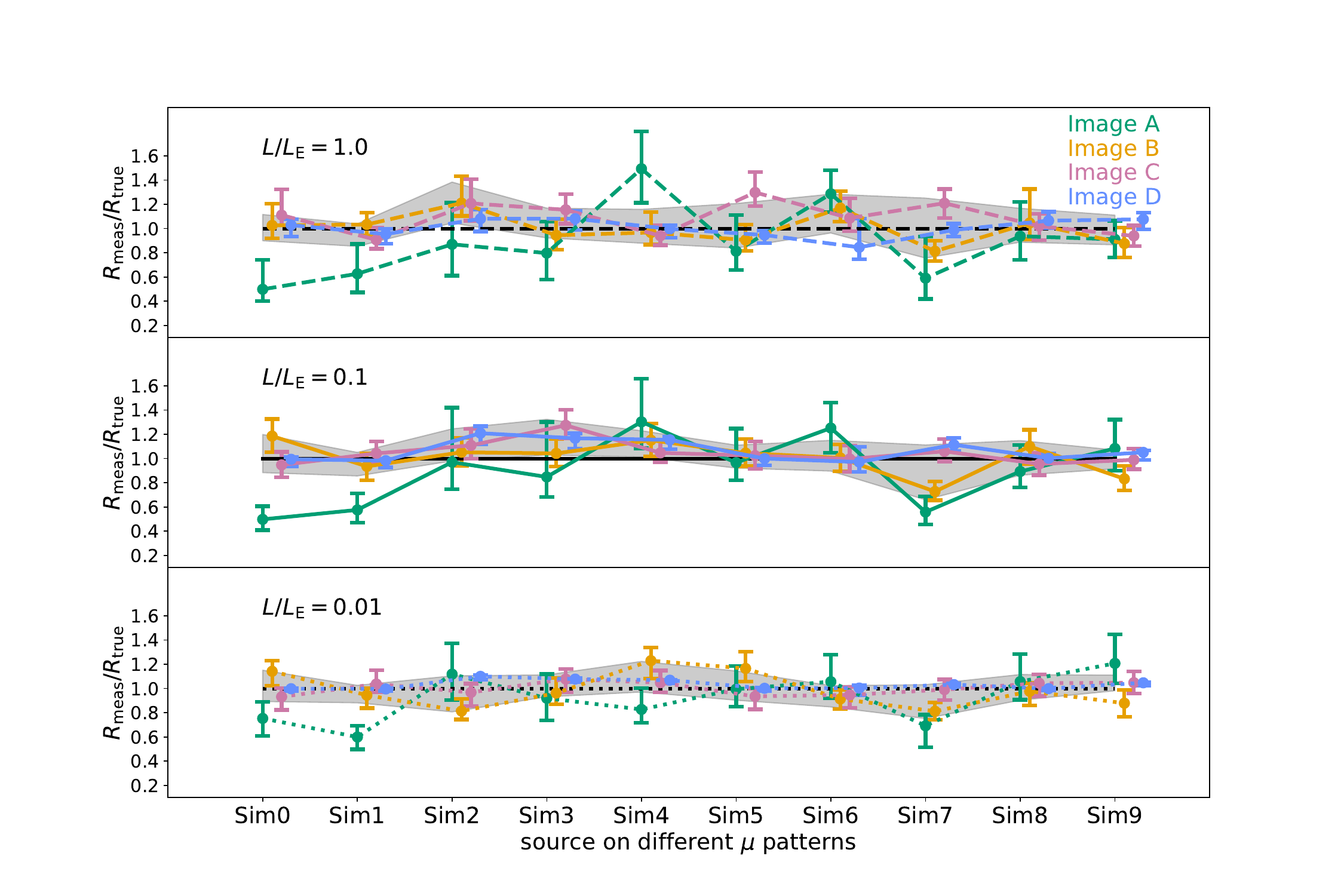}
\caption{
Same as \fref{fig:r0_imf} but now using the four images of J1131, as labeled on the top right corner. The template magnification maps are adopted using the IMF of $\Mstar=0.3,r=100$ and the marginalized result using four images are shown in the grey shade.}
\label{fig:r0_img}
\end{figure*}

\subsection{Simulation and method}
\label{subsec:method}

We provide a flowchart of our method in \fref{fig:method}. The flowchart can be split into three parts. We first start to simulate ``observed delays'', as illustrated in the top row of the figure and as explained below. We then analyse the simulated data by creating microlensing maps and by selecting possible locations where the microlensing time delays match the observed ones. This corresponds to the middle panel of the figure. Finally, we process the outputs of this selection and display them on the bottom row of the figure. Each of these steps are detailed below. 

We first simulate the data set by taking Rubin's filters, consisting in light curves in six bands, i.e. $u$, $g$, $r$, $i$, $z$, and $y$. 
The observed wavelengths for each filter are indicated on the top-right subplot of \fref{fig:method} for the specific case of image $A$ in J1131 that we take as a textbook case in this work. To do this, we put a disk on a given magnification pattern with size of $5\Rein$, as shown in the top-left subplot, and we generate the mean delay, $\DTmean$, in six Rubin-like bands. 
In this work, we disregard the peculiar velocity of the system, assuming static magnification patterns, which is achievable when delays can be measured within a short period \citep{MillonEtal20b}.
The (simulated) observed delay measurements are calculated by $\tau_{\lambda}=\DTmean-\DTmeanRef$, relative to a reference band ($\lambda_0$), labelled as red dots in the top-right subplot. These are our simulated data and we choose the bluest possible band as a reference, as this is where the accretion disk is the smallest. The relation of the resulting delays as a function of wavelength, with and without microlensing effect, are shown as green solid and blue dot-dashed lines, respectively. In this specific simulation, the delays become longer in presence of microlensing because there are more microcaustic falling on the outer parts of disk than on the inner parts, but of course the opposite case is possible, depending on where the microcaustics fall within the accretion disk. In practice, measuring such short delays is not trivial given the small source size, even for non-lensed quasars. Most traditional curve shifting techniques may underestimate source size by up to $\sim 30\%$, even in the absence of microlensing \citep{ChanEtal19}. 
This is because they often only consider a shift between the unmagnified light curves in each band and not the actual distortion caused by the convolution of the lamp-post disk profile by the light profile of the accretion disk. In the following, we will assume that the delays are measured without bias and we then illustrate how to use them to infer the accretion disk size.

Second, in the same spirit as reverberation mapping, we use a series of (correctly measured) multi-filter light curves for a given lensed image, and we perform a least-square fit following:
\be
\chi^2 = \sum_{\lambda}\frac{\left( \tau_{\lambda} + \DTmeanRef - \DTmean \right)^2}{\sigma_{\lambda}^2},
\label{eqn:chi2}
\ee
where $\sigma_{\lambda}$ is the rms precision of the measurement, that we assume here to be $0.1$~days. Unlike traditional reverberation mapping methods, there is no analytical relation between the delays at different wavelengths, since the relation $\DTmean\propto\lambda^{4/3}$ breaks down due to microlensing (See \sref{subsec:microlensing_delay} and \fref{fig:sim}). However, as shown in \fref{fig:sim}, we can predict relations between $\DTmean$ and $\Rsrc\propto\lambda^{4/3}$ for a given disk model and magnification pattern. To do so, we generate a large magnification map as a template, as shown in the middle row of \fref{fig:method}. Note that the input magnification pattern used to produce the simulated data is not taken from this template magnification map. We then place an accretion disk on this template magnification map and we seek for possible locations where the magnification may allow to match the observed delays $\tau_{\lambda}$. In practice, we first produce a series of relations between $\Rsrc$ and $\DTmean$ at all positions in the template map, by convolving with the corresponding accretion disk model of size $\Rsrc$. We then compare with the observed delays $\tau_\lambda$ with this series of predicted relations between $\Rsrc$ and $\DTmean$, using \eref{eqn:chi2}. Each fit provides us the best-fit of $\DTmeanRef$ at the specific location, and hence we can infer the corresponding $\RsrcRef$. This ``brute-force'' search is time consuming, since we need to scan over all positions on the template magnification map. Of course, the use of GPUs may speed this process up, but another way workaround is to reduce the number of fitting processes by preselecting the possible positions on the template map that reproduce the microlensing magnification. In the case of image A of J1131 this is $\mu=-4.28$, which also corresponds to the center of the input magnification pattern on the top-left subplot of \fref{fig:method}. 

In practice, this magnification is measurable on real data either using long light curves \citep{MillonEtal20} or flux anomalies \citep[e.g.,][]{MoreEtal17}. Here, we conservatively take the microlensing magnification with typical error bars of $25\%$. This allows us to reduce the number of locations on the map to $\lesssim10\%$ of what it would take without preselection, although relaxing this criterion is insensitive to the posterior size distribution. These plausible solutions are labelled as yellow areas in the Figure, which are the only places where we need to perform the least-square fitting. 

The last step is to collect those positions with small $\chi^2$ given a threshold ($\chi^2 < 1$ in this work), labelled as green dots on the template magnification map. In the bottom subplot, we draw the probability distribution of the source size estimation and highlight the mean and $1\sigma$ error bars the 50th, 16th and 84th percentiles of the distribution, respectively. The distribution shows that we are able to recover the source size within the $1\sigma$ error.

\subsection{Result}
\label{subsec:result}

With our analysis chain in hand, we now assign the disk source at 10 random positions of a magnification map to generate the observed delays (top panels in \fref{fig:sim}) and see whether we can recover the source size regardless of magnification pattern. Note that the input magnification patterns are produced using stellar masses with a mean mass of $\Mstar=0.3$ and assuming $r=100$ for the IMF. We test three source sizes by varying the quasar luminosity over a broad range, i.e. $\LLE=1.0$, $0.1$, and $0.01$, corresponding $\RsrcRef=0.277$, $0.129$, and $0.060$~light-day, respectively, according to \eref{eqn:source_size}. We then prepare a template magnification map in order to perform brute-force fitting. Hereafter we adopt 3 different IMFs which are common and accepted in microlensing works. The first two share the same mean stellar mass of $\Mstar=0.3$ but use $r=100$ and $r=1$. The last one has $\Mstar=1.0$ and $r=1$. 
Each magnification template map has $20\Rein$ on-a-side. 
Although here we only present the result with one realization of the template for each IMF, the test for more realizations is described in \aref{sec:more_test}.

The distribution of size measurements for each simulation is presented in \fref{fig:r0_imf}. Each data point and error bar correspond the 50th, 16th and 84th percentiles of the probability distribution, as shown in the bottom panel of \fref{fig:method}. We then compare with the true value and therefore show the ratio between the measurement and true value $R_{\rm meas}/R_{\rm true}$. It is immediately striking that the measurement is insensitive to choice of the IMFs in the template microlensing map.
This is particularly important since so far determining the IMF of lensing galaxy towards the line of sight to a lensed quasar image is not possible. Still, we are able to recover the disk size within a factor of $2$, regardless of the input size. 

A nice advantages of lensed quasar over non-lensed ones where reverberation mapping is usually performed, is that we have two or even four sets of multi-band light curves for each object. In the ideal case of fours images, this allows us to measure the source size four times simultaneously, as we illustrate in \fref{fig:r0_img}. Here we only employ the template magnification maps with $\Mstar=0.3$ and $r=100$, since \fref{fig:r0_imf} shows that the measurement is insensitive to the IMFs. After marginalizing over all fours measurements, as shown in the grey shade, we are able to recover the source within the error $\sim 20\%$. When a source is smaller, the error is also slightly smaller, since the microlensing time delay is closer in amplitude to the geometric delay. The uncertainty of measurement of image $A$ is larger than other lensed images, as a result of the stronger microlensing effect. Of course, it is also possible to apply this method to doubly lensed quasars.
Although this gives only access to two sets of multi-band light curves, microlensing is in general less effective in the brighter image of doubles.

For clarity, all the above study is carried out for a face-on accretion disk. Other disk configurations with inclinations and position angles can also be considered and are explored in \aref{sec:more_test}.
In this work, we choose the exceptional condition (rms 0.1~days) to demonstrate this method, and achieve the comparable source-size measurement with that from other techniques.
We note that high accuracy and precision of delay is like to be necessary to this method, since the microlensing time delays between filters are in general very short.
The requirement of the delay measurement is beyond the scope of this work.

\section{Conclusion and discussion}
\label{sec:conclusion}

TK18 first introduced the notion of microlensing time delay and argued that time delay measurements in lensed quasars can be biased, due to different parts of the accretion disk being microlensed in different ways. Since the effect depends on the source size and since the source size is a chromatic quantity, we propose to measure multi-band time delays in lensed quasars and to use them to measure the size the accretion disk under the assumption of the lamp-post model and given a model for the accretion disk light profile. 

We investigate the microlensing time delay using different IMFs taking the well-studied lensed quasar J1131 as an example and show that the magnification probability distribution is insensitive to the IMF in the lensing galaxy. 
The probability distribution of microlensing time delay is wider when decreasing the mean stellar mass of the microlenses.
This trend follows accordingly as increasing the source size, as the microlensing analysis can be scaled by the mean mass of microlenses.
Still, for typical mean stellar masses $\Mstar$ between $0.3$ and $1.0~\Msun$, the distributions for microlensing time delays are nearly indistinguishable. 
This makes it possible to measure disk sizes even under the assumption of poorly known IMF or mean stellar masses, contrary to traditional microlensing methods, where the source size is very degenerate with the properties assumed for the microlenses.

We propose a new method to measure the disk size of quasars, based on the continuum reverberation mapping method but applied to strongly lensed quasars, assuming the static magnification pattern. To illustrate the method, we assume Rubin-like light curves providing time delays in 6 filters with each an rms precision of 0.1~days and no bias. 
Assuming that such data can actually be gathered in practice, we find that 
\begin{itemize}
\item Using the brute-force search on a template magnification map, we are able to recover the disk size within a factor of $2$, with the multi-filter light curves of only one lensed image. 
\item This method is insensitive to the IMFs. 
\item After marginalizing all measurements of multiple lensed images, we can achieve an unbiased measurement within error $\sim20\%$
\end{itemize}

The present work sets the basis for a new way of using microlensing time delay in multi-band light curves for astrophysical purpose. 
We focus here on how to measure the size of quasar accretion disks, but our future work will also consider the joint measurements of the cosmological time delay and microlensing time delay, allowing us both to mitigate the impact of microlensing time delay on $H_0$ measurements and to use it to measure quasar accretion disks, at the same time. 
This is particularly relevant in the era of large time-domain and multi-band surveys like the Rubin Observatory Legacy Survey of Space and Time (LSST).


\section*{Acknowledgements}
We thank P.~Schechter for the useful discussion and the referee for the comments.
This work is supported by the Swiss National Science Foundation (SNSF) and by the European Research Council (ERC) under the European Union’s Horizon 2020 research and innovation program (COSMICLENS: grant agreement No 787886).


\bibliographystyle{aa}
\bibliography{reference}

\begin{thebibliography}{53}
\expandafter\ifx\csname natexlab\endcsname\relax\def\natexlab#1{#1}\fi

\bibitem[{{Abramowicz} {et~al.}(1988){Abramowicz}, {Czerny}, {Lasota}, \&
  {Szuszkiewicz}}]{AbramowiczEtal88}
{Abramowicz}, M.~A., {Czerny}, B., {Lasota}, J.~P., \& {Szuszkiewicz}, E. 1988,
  \apj, 332, 646

\bibitem[{{Birrer} {et~al.}(2019){Birrer}, {Treu}, {Rusu}, {Bonvin},
  {Fassnacht}, {Chan}, {Agnello}, {Shajib}, {Chen}, {Auger}, {Courbin},
  {Hilbert}, {Sluse}, {Suyu}, {Wong}, {Marshall}, {Lemaux}, \&
  {Meylan}}]{BirrerEtal19}
{Birrer}, S., {Treu}, T., {Rusu}, C.~E., {et~al.} 2019, \mnras, 484, 4726

\bibitem[{{Bonvin} {et~al.}(2018){Bonvin}, {Chan}, {Millon}, {Rojas},
  {Courbin}, {Chen}, {Fassnacht}, {Paic}, {Tewes}, {Chao}, {Chijani}, {Gilman},
  {Gilmore}, {Williams}, {Buckley-Geer}, {Frieman}, {Marshall}, {Suyu}, {Treu},
  {Hempel}, {Kim}, {Lachaume}, {Rabus}, {Anguita}, {Meylan}, {Motta}, \&
  {Magain}}]{BonvinEtal18}
{Bonvin}, V., {Chan}, J.~H.~H., {Millon}, M., {et~al.} 2018, \aap, 616, A183

\bibitem[{{Bonvin} {et~al.}(2019{\natexlab{a}}){Bonvin}, {Millon}, {Chan},
  {Courbin}, {Rusu}, {Sluse}, {Suyu}, {Wong}, {Fassnacht}, {Marshall}, {Treu},
  {Buckley-Geer}, {Frieman}, {Hempel}, {Kim}, {Lachaume}, {Rabus}, {Chao},
  {Chijani}, {Gilman}, {Gilmore}, {Rojas}, {Williams}, {Anguita}, {Kochanek},
  {Morgan}, {Motta}, {Tewes}, \& {Meylan}}]{BonvinEtal19b}
{Bonvin}, V., {Millon}, M., {Chan}, J.~H.~H., {et~al.} 2019{\natexlab{a}},
  \aap, 629, A97

\bibitem[{{Bonvin} {et~al.}(2019{\natexlab{b}}){Bonvin}, {Tihhonova}, {Millon},
  {Chan}, {Savary}, {Huber}, \& {Courbin}}]{BonvinEtal19a}
{Bonvin}, V., {Tihhonova}, O., {Millon}, M., {et~al.} 2019{\natexlab{b}}, \aap,
  621, A55

\bibitem[{{Cackett} {et~al.}(2007){Cackett}, {Horne}, \&
  {Winkler}}]{CackettEtal07}
{Cackett}, E.~M., {Horne}, K., \& {Winkler}, H. 2007, \mnras, 380, 669

\bibitem[{{Chabrier}(2003)}]{Chabrier03}
{Chabrier}, G. 2003, \pasp, 115, 763

\bibitem[{{Chan} {et~al.}(2019){Chan}, {Millon}, {Bonvin}, \&
  {Courbin}}]{ChanEtal19}
{Chan}, J.~H.~H., {Millon}, M., {Bonvin}, V., \& {Courbin}, F. 2019, arXiv
  e-prints, arXiv:1909.08638

\bibitem[{{Chen} {et~al.}(2018){Chen}, {Chan}, {Bonvin}, {Fassnacht}, {Rojas},
  {Millon}, {Courbin}, {Suyu}, {Wong}, {Sluse}, {Treu}, {Shajib}, {Hsueh},
  {Lagattuta}, {Koopmans}, {Vegetti}, \& {McKean}}]{ChenEtal18}
{Chen}, G. C.~F., {Chan}, J. H.~H., {Bonvin}, V., {et~al.} 2018, \mnras, 481,
  1115

\bibitem[{{Chen} {et~al.}(2019){Chen}, {Fassnacht}, {Suyu}, {Rusu}, {Chan},
  {Wong}, {Auger}, {Hilbert}, {Bonvin}, {Birrer}, {Millon}, {Koopmans},
  {Lagattuta}, {McKean}, {Vegetti}, {Courbin}, {Ding}, {Halkola}, {Jee},
  {Shajib}, {Sluse}, {Sonnenfeld}, \& {Treu}}]{ChenEtal19}
{Chen}, G. C.~F., {Fassnacht}, C.~D., {Suyu}, S.~H., {et~al.} 2019, \mnras,
  490, 1743

\bibitem[{{Collier} {et~al.}(1998){Collier}, {Horne}, {Kaspi}, {Netzer},
  {Peterson}, {Wanders}, {Alexander}, {Bertram}, {Comastri}, {Gaskell},
  {Malkov}, {Maoz}, {Mignoli}, {Pogge}, {Pronik}, {Sergeev}, {Snedden},
  {Stirpe}, {Bochkarev}, {Burenkov}, {Shapovalova}, \&
  {Wagner}}]{CollierEtal98}
{Collier}, S.~J., {Horne}, K., {Kaspi}, S., {et~al.} 1998, \apj, 500, 162

\bibitem[{{Cornachione} \& {Morgan}(2020)}]{Cornachione&Morgan20}
{Cornachione}, M.~A. \& {Morgan}, C.~W. 2020, arXiv e-prints, arXiv:2006.07243

\bibitem[{Cornachione {et~al.}(2020)Cornachione, Morgan, Millon, Bentz,
  Courbin, Bonvin, \& Falco}]{CornachioneEtal20}
Cornachione, M.~A., Morgan, C.~W., Millon, M., {et~al.} 2020, The Astrophysical
  Journal, 895, 125

\bibitem[{{Dai} {et~al.}(2010){Dai}, {Kochanek}, {Chartas}, {Koz{\l}owski},
  {Morgan}, {Garmire}, \& {Agol}}]{DaiEtal10}
{Dai}, X., {Kochanek}, C.~S., {Chartas}, G., {et~al.} 2010, \apj, 709, 278

\bibitem[{{Dalal} \& {Kochanek}(2002)}]{DalalKochanek02}
{Dalal}, N. \& {Kochanek}, C.~S. 2002, \apj, 572, 25

\bibitem[{{Fausnaugh} {et~al.}(2016){Fausnaugh}, {Denney}, {Barth}, {Bentz},
  {Bottorff}, {Carini}, {Croxall}, {De Rosa}, {Goad}, {Horne}, {Joner},
  {Kaspi}, {Kim}, {Klimanov}, {Kochanek}, {Leonard}, {Netzer}, {Peterson},
  {Schn{\"u}lle}, {Sergeev}, {Vestergaard}, {Zheng}, {Zu}, {Anderson},
  {Ar{\'e}valo}, {Bazhaw}, {Borman}, {Boroson}, {Brandt}, {Breeveld}, {Brewer},
  {Cackett}, {Crenshaw}, {Dalla Bont{\`a}}, {De Lorenzo-C{\'a}ceres},
  {Dietrich}, {Edelson}, {Efimova}, {Ely}, {Evans}, {Filippenko}, {Flatland},
  {Gehrels}, {Geier}, {Gelbord}, {Gonzalez}, {Gorjian}, {Grier}, {Grupe},
  {Hall}, {Hicks}, {Horenstein}, {Hutchison}, {Im}, {Jensen}, {Jones},
  {Kaastra}, {Kelly}, {Kennea}, {Kim}, {Korista}, {Kriss}, {Lee}, {Lira},
  {MacInnis}, {Manne-Nicholas}, {Mathur}, {McHardy}, {Montouri}, {Musso},
  {Nazarov}, {Norris}, {Nousek}, {Okhmat}, {Pancoast}, {Papadakis}, {Parks},
  {Pei}, {Pogge}, {Pott}, {Rafter}, {Rix}, {Saylor}, {Schimoia}, {Siegel},
  {Spencer}, {Starkey}, {Sung}, {Teems}, {Treu}, {Turner}, {Uttley},
  {Villforth}, {Weiss}, {Woo}, {Yan}, \& {Young}}]{FausnaughEtal16}
{Fausnaugh}, M.~M., {Denney}, K.~D., {Barth}, A.~J., {et~al.} 2016, \apj, 821,
  56

\bibitem[{{Harvey} {et~al.}(2020){Harvey}, {Valkenburg}, {Tamone}, {Boyarsky},
  {Courbin}, \& {Lovell}}]{Harvey2020}
{Harvey}, D., {Valkenburg}, W., {Tamone}, A., {et~al.} 2020, \mnras, 491, 4247

\bibitem[{{Ivezi{\'c}} {et~al.}(2019){Ivezi{\'c}}, {Kahn}, {Tyson}, {Abel},
  {Acosta}, {Allsman}, {Alonso}, {AlSayyad}, {Anderson}, {Andrew}, \&
  et~al.}]{IvezicEtal19}
{Ivezi{\'c}}, {\v{Z}}., {Kahn}, S.~M., {Tyson}, J.~A., {et~al.} 2019, \apj,
  873, 111

\bibitem[{{Jiang} {et~al.}(2017){Jiang}, {Green}, {Greene}, {Morganson},
  {Shen}, {Pancoast}, {MacLeod}, {Anderson}, {Brandt}, {Grier}, {Rix}, {Ruan},
  {Protopapas}, {Scott}, {Burgett}, {Hodapp}, {Huber}, {Kaiser}, {Kudritzki},
  {Magnier}, {Metcalfe}, {Tonry}, {Wainscoat}, \& {Waters}}]{JiangEtal17}
{Jiang}, Y.-F., {Green}, P.~J., {Greene}, J.~E., {et~al.} 2017, \apj, 836, 186

\bibitem[{{Jim{\'e}nez-Vicente} \&
  {Mediavilla}(2019)}]{Jimnez-Vicente&Mediavilla19}
{Jim{\'e}nez-Vicente}, J. \& {Mediavilla}, E. 2019, \apj, 885, 75

\bibitem[{{Kayser} {et~al.}(1986){Kayser}, {Refsdal}, \&
  {Stabell}}]{KayserEtal86}
{Kayser}, R., {Refsdal}, S., \& {Stabell}, R. 1986, \aap, 166, 36

\bibitem[{{Kochanek}(2004)}]{Kochanek04}
{Kochanek}, C.~S. 2004, \apj, 605, 58

\bibitem[{{Krolik} {et~al.}(1991){Krolik}, {Horne}, {Kallman}, {Malkan},
  {Edelson}, \& {Kriss}}]{KrolikEtal91}
{Krolik}, J.~H., {Horne}, K., {Kallman}, T.~R., {et~al.} 1991, \apj, 371, 541

\bibitem[{{Lewis} \& {Irwin}(1995)}]{Lewis&Irwin95}
{Lewis}, G.~F. \& {Irwin}, M.~J. 1995, \mnras, 276, 103

\bibitem[{{Mediavilla} {et~al.}(2015){Mediavilla}, {Jimenez-Vicente},
  {Mu{\~n}oz}, {Mediavilla}, \& {Ariza}}]{MediavillaEtal15}
{Mediavilla}, E., {Jimenez-Vicente}, J., {Mu{\~n}oz}, J.~A., {Mediavilla}, T.,
  \& {Ariza}, O. 2015, \apj, 798, 138

\bibitem[{{Millon} {et~al.}(2020{\natexlab{a}}){Millon}, {Courbin}, {Bonvin},
  {Buckley-Geer}, {Fassnacht}, {Frieman}, {Marshall}, {Suyu}, {Treu},
  {Anguita}, {Motta}, {Agnello}, {Chan}, {C. -Y Chao}, {Chijani}, {Gilman},
  {Gilmore}, {Lemon}, {Lucey}, {Melo}, {Paic}, {Rojas}, {Sluse}, {Williams},
  {Hempel}, {Kim}, {Lachaume}, \& {Rabus}}]{MillonEtal20b}
{Millon}, M., {Courbin}, F., {Bonvin}, V., {et~al.} 2020{\natexlab{a}}, arXiv
  e-prints, arXiv:2006.10066

\bibitem[{{Millon} {et~al.}(2020{\natexlab{b}}){Millon}, {Courbin}, {Bonvin},
  {Paic}, {Meylan}, {Tewes}, {Sluse}, {Magain}, {Chan}, {Galan}, {Joseph},
  {Lemon}, {Tihhonova}, {Anderson}, {Marmier}, {Chazelas}, {Lendl}, {Triaud},
  \& {Wyttenbach}}]{MillonEtal20}
{Millon}, M., {Courbin}, F., {Bonvin}, V., {et~al.} 2020{\natexlab{b}}, arXiv
  e-prints, arXiv:2002.05736

\bibitem[{{Millon} {et~al.}(2019){Millon}, {Galan}, {Courbin}, {Treu}, {Suyu},
  {Ding}, {Birrer}, {Chen}, {Shajib}, {Wong}, {Agnello}, {Auger},
  {Buckley-Geer}, {Chan}, {Collett}, {Fassnacht}, {Hilbert}, {Koopmans},
  {Motta}, {Mukherjee}, {Rusu}, {Sluse}, {Sonnenfeld}, {Spiniello}, \& {Van de
  Vyvere}}]{MillonEtal19}
{Millon}, M., {Galan}, A., {Courbin}, F., {et~al.} 2019, arXiv e-prints,
  arXiv:1912.08027

\bibitem[{{More} {et~al.}(2017){More}, {Suyu}, {Oguri}, {More}, \&
  {Lee}}]{MoreEtal17}
{More}, A., {Suyu}, S.~H., {Oguri}, M., {More}, S., \& {Lee}, C.-H. 2017,
  \apjl, 835, L25

\bibitem[{{Morgan} {et~al.}(2018){Morgan}, {Hyer}, {Bonvin}, {Mosquera},
  {Cornachione}, {Courbin}, {Kochanek}, \& {Falco}}]{MorganEtal18}
{Morgan}, C.~W., {Hyer}, G.~E., {Bonvin}, V., {et~al.} 2018, \apj, 869, 106

\bibitem[{{Morgan} {et~al.}(2010){Morgan}, {Kochanek}, {Morgan}, \&
  {Falco}}]{MorganEtal10}
{Morgan}, C.~W., {Kochanek}, C.~S., {Morgan}, N.~D., \& {Falco}, E.~E. 2010,
  \apj, 712, 1129

\bibitem[{{Mudd} {et~al.}(2018){Mudd}, {Martini}, {Zu}, {Kochanek}, {Peterson},
  {Kessler}, {Davis}, {Hoormann}, {King}, {Lidman}, {Sommer}, {Tucker},
  {Asorey}, {Hinton}, {Glazebrook}, {Kuehn}, {Lewis}, {Macaulay}, {Moeller},
  {O'Neill}, {Zhang}, {Abbott}, {Abdalla}, {Allam}, {Banerji},
  {Benoit-L{\'e}vy}, {Bertin}, {Brooks}, {Carnero Rosell}, {Carollo}, {Carrasco
  Kind}, {Carretero}, {Cunha}, {D'Andrea}, {da Costa}, {Davis}, {Desai},
  {Doel}, {Fosalba}, {Garc{\'{\i}}a-Bellido}, {Gaztanaga}, {Gerdes}, {Gruen},
  {Gruendl}, {Gschwend}, {Gutierrez}, {Hartley}, {Honscheid}, {James},
  {Kuhlmann}, {Kuropatkin}, {Lima}, {Maia}, {Marshall}, {McMahon}, {Menanteau},
  {Miquel}, {Plazas}, {Romer}, {Sanchez}, {Schindler}, {Schubnell}, {Smith},
  {Smith}, {Soares-Santos}, {Sobreira}, {Suchyta}, {Swanson}, {Tarle},
  {Thomas}, {Tucker}, {Walker}, \& {DES Collaboration}}]{MuddEtal18}
{Mudd}, D., {Martini}, P., {Zu}, Y., {et~al.} 2018, \apj, 862, 123

\bibitem[{{Oguri} {et~al.}(2014){Oguri}, {Rusu}, \& {Falco}}]{OguriEtal14}
{Oguri}, M., {Rusu}, C.~E., \& {Falco}, E.~E. 2014, \mnras, 439, 2494

\bibitem[{{Refsdal}(1964)}]{Refsdal64}
{Refsdal}, S. 1964, \mnras, 128, 307

\bibitem[{Rojas {et~al.}(2014)Rojas, Motta, Mediavilla, Falco,
  Jiménez-Vicente, \& Muñoz}]{RojasEtal14}
Rojas, K., Motta, V., Mediavilla, E., {et~al.} 2014, The Astrophysical Journal,
  797, 61

\bibitem[{Rojas {et~al.}(2020)Rojas, Motta, Mediavilla, Jiménez-Vicente,
  Falco, \& Fian}]{RojasEtal20}
Rojas, K., Motta, V., Mediavilla, E., {et~al.} 2020, The Astrophysical Journal,
  890, 3

\bibitem[{{Schechter} \& {Wambsganss}(2002)}]{Schechter&Wambsganss02}
{Schechter}, P.~L. \& {Wambsganss}, J. 2002, \apj, 580, 685

\bibitem[{{Schechter} {et~al.}(2004){Schechter}, {Wambsganss}, \&
  {Lewis}}]{SchechterEtal04}
{Schechter}, P.~L., {Wambsganss}, J., \& {Lewis}, G.~F. 2004, \apj, 613, 77

\bibitem[{{Shajib} {et~al.}(2019){Shajib}, {Birrer}, {Treu}, {Agnello},
  {Buckley-Geer}, {Chan}, {Christensen}, {Lemon}, {Lin}, {Millon}, {Poh},
  {Rusu}, {Sluse}, {Spiniello}, {Chen}, {Collett}, {Courbin}, {Fassnacht},
  {Frieman}, {Galan}, {Gilman}, {More}, {Anguita}, {Auger}, {Bonvin},
  {McMahon}, {Meylan}, {Wong}, {Abbott}, {Annis}, {Avila}, {Bechtol}, {Brooks},
  {Brout}, {Burke}, {Carnero Rosell}, {Carrasco Kind}, {Carretero},
  {Castander}, {Costanzi}, {da Costa}, {De Vicente}, {Desai}, {Dietrich},
  {Doel}, {Drlica-Wagner}, {Evrard}, {Finley}, {Flaugher}, {Fosalba},
  {Garc{\'\i}a-Bellido}, {Gerdes}, {Gruen}, {Gruendl}, {Gschwend}, {Gutierrez},
  {Hollowood}, {Honscheid}, {Huterer}, {James}, {Jeltema}, {Krause},
  {Kuropatkin}, {Li}, {Lima}, {MacCrann}, {Maia}, {Marshall}, {Melchior},
  {Miquel}, {Ogando}, {Palmese}, {Paz-Chinch{\'o}n}, {Plazas}, {Romer},
  {Roodman}, {Sako}, {Sanchez}, {Santiago}, {Scarpine}, {Schubnell}, {Scolnic},
  {Serrano}, {Sevilla-Noarbe}, {Smith}, {Soares-Santos}, {Suchyta}, {Tarle},
  {Thomas}, {Walker}, \& {Zhang}}]{ShajibEtal19}
{Shajib}, A.~J., {Birrer}, S., {Treu}, T., {et~al.} 2019, arXiv e-prints,
  arXiv:1910.06306

\bibitem[{{Shakura} \& {Sunyaev}(1973)}]{Shakura&Sunyaev73}
{Shakura}, N.~I. \& {Sunyaev}, R.~A. 1973, \aap, 24, 337

\bibitem[{{Sluse} {et~al.}(2003){Sluse}, {Surdej}, {Claeskens},
  {Hutsem{\'e}kers}, {Jean}, {Courbin}, {Nakos}, {Billeres}, \&
  {Khmil}}]{SluseEtal03}
{Sluse}, D., {Surdej}, J., {Claeskens}, J.~F., {et~al.} 2003, \aap, 406, L43

\bibitem[{{Starkey} {et~al.}(2016){Starkey}, {Horne}, \&
  {Villforth}}]{StarkeyEtal16}
{Starkey}, D.~A., {Horne}, K., \& {Villforth}, C. 2016, \mnras, 456, 1960

\bibitem[{{Suyu} {et~al.}(2017){Suyu}, {Bonvin}, {Courbin}, {Fassnacht},
  {Rusu}, {Sluse}, {Treu}, {Wong}, {Auger}, {Ding}, {Hilbert}, {Marshall},
  {Rumbaugh}, {Sonnenfeld}, {Tewes}, {Tihhonova}, {Agnello}, {Blandford},
  {Chen}, {Collett}, {Koopmans}, {Liao}, {Meylan}, \& {Spiniello}}]{SuyuEtal17}
{Suyu}, S.~H., {Bonvin}, V., {Courbin}, F., {et~al.} 2017, \mnras, 468, 2590

\bibitem[{{Suyu} {et~al.}(2012){Suyu}, {Hensel}, {McKean}, {Fassnacht}, {Treu},
  {Halkola}, {Norbury}, {Jackson}, {Schneider}, {Thompson}, {Auger},
  {Koopmans}, \& {Matthews}}]{SuyuEtal12}
{Suyu}, S.~H., {Hensel}, S.~W., {McKean}, J.~P., {et~al.} 2012, \apj, 750, 10

\bibitem[{{Suyu} {et~al.}(2010){Suyu}, {Marshall}, {Auger}, {Hilbert},
  {Blandford}, {Koopmans}, {Fassnacht}, \& {Treu}}]{SuyuEtal10}
{Suyu}, S.~H., {Marshall}, P.~J., {Auger}, M.~W., {et~al.} 2010, \apj, 711, 201

\bibitem[{{Tie} \& {Kochanek}(2018)}]{Tie&Kochanek18}
{Tie}, S.~S. \& {Kochanek}, C.~S. 2018, \mnras, 473, 80

\bibitem[{{Vernardos} \& {Fluke}(2013)}]{Vernardos&Fluke13}
{Vernardos}, G. \& {Fluke}, C.~J. 2013, \mnras, 434, 832

\bibitem[{{Vernardos} \& {Fluke}(2014)}]{Vernardos&Fluke14}
{Vernardos}, G. \& {Fluke}, C.~J. 2014, Astronomy and Computing, 6, 1

\bibitem[{{Wambsganss}(1992)}]{Wambsganss92}
{Wambsganss}, J. 1992, \apj, 386, 19

\bibitem[{{Wanders} {et~al.}(1997){Wanders}, {Peterson}, {Alloin}, {Ayres},
  {Clavel}, {Crenshaw}, {Horne}, {Kriss}, {Krolik}, {Malkan}, {Netzer},
  {O'Brien}, {Reichert}, {Rodr{\'\i}guez-Pascual}, {Wamsteker}, {Alexand er},
  {Anderson}, {Benitez}, {Bochkarev}, {Burenkov}, {Cheng}, {Collier},
  {Comastri}, {Dietrich}, {Dultzin-Hacyan}, {Espey}, {Filippenko}, {Gaskell},
  {George}, {Goad}, {Ho}, {Kaspi}, {Kollatschny}, {Korista}, {Laor},
  {MacAlpine}, {Mignoli}, {Morris}, {Nand ra}, {Penton}, {Pogge}, {Ptak},
  {Rodr{\'\i}guez-Espinoza}, {Santos-Lle{\'o}}, {Shapovalova}, {Shull},
  {Snedden}, {Sparke}, {Stirpe}, {Sun}, {Turner}, {Ulrich}, {Wang}, {Wei},
  {Welsh}, {Xue}, \& {Zou}}]{WandersEtal97}
{Wanders}, I., {Peterson}, B.~M., {Alloin}, D., {et~al.} 1997, \apjs, 113, 69

\bibitem[{{Wong} {et~al.}(2019){Wong}, {Suyu}, {Chen}, {Rusu}, {Millon},
  {Sluse}, {Bonvin}, {Fassnacht}, {Taubenberger}, {Auger}, {Birrer}, {Chan},
  {Courbin}, {Hilbert}, {Tihhonova}, {Treu}, {Agnello}, {Ding}, {Jee},
  {Komatsu}, {Shajib}, {Sonnenfeld}, {Bland ford}, {Koopmans}, {Marshall}, \&
  {Meylan}}]{WongEtal19}
{Wong}, K.~C., {Suyu}, S.~H., {Chen}, G. C.~F., {et~al.} 2019, arXiv e-prints,
  arXiv:1907.04869

\bibitem[{{Wyithe} \& {Turner}(2001)}]{Wyithe&Turner01}
{Wyithe}, J.~S.~B. \& {Turner}, E.~L. 2001, \mnras, 320, 21

\bibitem[{{Yu} {et~al.}(2018){Yu}, {Martini}, {Davis}, {Gruendl}, {Hoormann},
  {Kochanek}, {Lidman}, {Mudd}, {Peterson}, {Wester}, {Allam}, {Annis},
  {Asorey}, {Avila}, {Banerji}, {Bertin}, {Brooks}, {Buckley-Geer}, {Calcino},
  {Carnero Rosell}, {Carollo}, {Carrasco Kind}, {Carretero}, {Cunha},
  {D'Andrea}, {da Costa}, {De Vicente}, {Desai}, {Diehl}, {Doel}, {Eifler},
  {Flaugher}, {Fosalba}, {Frieman}, {Garc{\'{\i}}a-Bellido}, {Gaztanaga},
  {Glazebrook}, {Gruen}, {Gschwend}, {Gutierrez}, {Hartley}, {Hinton},
  {Hollowood}, {Honscheid}, {Hoyle}, {James}, {Kim}, {Krause}, {Kuehn},
  {Kuropatkin}, {Lewis}, {Lima}, {Macaulay}, {Maia}, {Marshall}, {Menanteau},
  {Miquel}, {M{\"o}ller}, {Plazas}, {Romer}, {Sanchez}, {Scarpine},
  {Schubnell}, {Serrano}, {Smith}, {Smith}, {Soares-Santos}, {Sobreira},
  {Suchyta}, {Swann}, {Swanson}, {Tarle}, {Tucker}, {Tucker}, \&
  {Vikram}}]{YuEtal18}
{Yu}, Z., {Martini}, P., {Davis}, T.~M., {et~al.} 2018, ArXiv e-prints
  [\eprint[arXiv]{1811.03638}]

\end{thebibliography}

\appendix
\section{Additional tests}
\label{sec:more_test}

\begin{figure}
\centering
\includegraphics[scale=0.45]{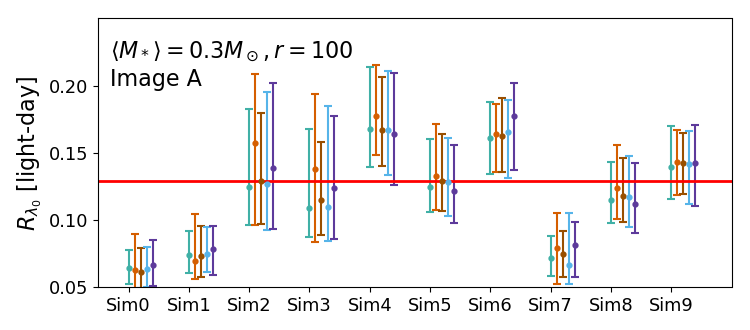}
\caption{
Source measurement of Image $A$ using different template magnification maps. The source is placed on $10$ magnification patterns labeled from Sim0 to Sim9.
For each simulation, we employ five realizations of templates to measure the source size, as labeled with different colors.
The true value is labeled with the red line.
}
\label{fig:r0_sn}
\end{figure}

In this section, we present the test on the different realizations of template magnification map and the test on the tilted disk.
From each simulation, we measure the source size with five different realizations of template magnification maps with $20\Rein$ on-a-side, as labelled with different colors.
Each map is generated using the IMF of $\Mstar=0.3$ and $r=100$.
\fref{fig:r0_sn}, same as \fref{fig:r0_imf}, shows that size measurements from different realizations agree well. 
This test also demonstrate that size of the template maps is large enough for this method.

\begin{figure}
\centering
\includegraphics[scale=0.45]{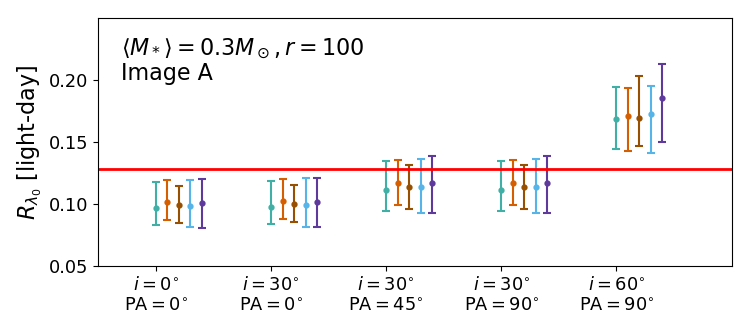}
\caption{
Source measurement of Image $A$ with different inclinations ($i$) and position angles (PA) of disk configuration.
For each simulation, we employ five realizations of template magnification map to measure the source size.
The true value is labeled with the red line.
}
\label{fig:r0_incl}
\end{figure}

In this work, we mainly present our result with face-on disks.
We further explore the tilted disks with inclination ($i$) and position angle (PA).
The inclination angle is defined as the tilted degree along the line of sight, with $i=0^{\circ}$ corresponding to the disk lying in the plane of the sky.
The position angle determines an angle between the long axis of the tilted disk and the caustic structures of the magnification maps.
Although the observed delays are simulated using a tilted disk, we still employ our method with face-on disk ($i=0^{\circ}$ and $\text{PA}=0^{\circ}$) to measure the source size.
The result is shown in \fref{fig:r0_incl}.

Larger inclination angle or position angle give us slightly larger measurements, resulting from the alignment between the disk and caustics.
This result agrees with TK18, which found that the microlensing time delay has minor dependence of the inclination of the disk and its orientation relative to the caustic networks.
Therefore, the assumption of the face-on disk is sufficient to recover the input disk size.
%


\end{document}